# An Investigation of Natural Gas as a Substitute for Diesel in Heavy Duty Trucks and Associated Considerations

Muneer Mohammad, Student Member, IEEE, *Mehrdad Ehsani,* Fellow, IEEE

*Abstract*— **In this paper, applicability of natural gas fuel for transportation as compared to diesel are investigated. This study investigates a promising technology for the heavy duty truck sector of transportation as a target for conversion from diesel to natural gas. The supply of natural gas is limited so we also verify the available domestic supply quantities both before and after a fleet conversion. This paper concludes with an economic discussion regarding Jevon's paradox and the fungibility of natural gas as compared to that of oil. In order to determine if natural gas can replace diesel for the country's heavy duty truck transportation needs, the energy equivalent and efficiencies of natural gas alternatives should be compared to diesel. There are two alternatives for using natural gas as a replacement for diesel: compressed natural gas (CNG) and liquefied natural gas (LNG).**

## I. INTRODUCTION

For several decades, natural gas has provided clean power to thousands of households and business nationwide. Natural gas is a promising clean fuel source. The use of clean alternative fuels opens new opportunities to consumers to address concerns about fuel costs, emissions, and provides them with new fuel supply choices. Research and development, along with several projects, have been directed by different organizations and laboratories in order to study the benefits of using Liquefied Natural Gas (LNG) as a fuel in heavy-duty vehicles. In [20] a cooperative on-road development project was developed. This project led to a production of a heavy-duty natural gas engine by Detroit Diesel Corporation. This project demonstrated that LNG can be a viable fuel for heavy-duty trucks and that it offers a very promising alternative to diesel fuel. The authors of [21] suggest the use of LNG as a clean alternative fuel, based on its merits of safety, high efficiency, low cost and environmental friendly fuel.

Natural gas is a promising fuel source. It is used already in the United States and around the world. The uses for natural gas include residential and commercial heating, industrial applications like reformation and most importantly as a fuel source for electrical generation [8]. Conventional natural gas has been a plentiful resource in places like Qatar, Russia, and the United States. With advances in technology, unconventional natural gas has been shown to be even more plentiful in even more places around the world. This increase in the world's supply of natural gas combined with increased demand on other energy sources has prompted calls for a broadening use of natural gas, especially in transportation.

In this paper, we analyze this call for natural gas use in transportation. We first present a better understanding of the energy comparability of natural gas to diesel. We then identify the heavy duty truck sector of transportation as a target for conversion from diesel to natural gas and the factors associated with this conversion. The supply of natural gas is limited, so we also verify the available domestic supply both before and after a fleet conversion. Also of interest is the increased consumption due to conversion on the transport capacity of natural gas. Environmentally, a switch from diesel to natural gas has prompted discussion of possible reduction in $CO_2$ emissions. This will also be analyzed in this paper. We will end with an economic discussion regarding Jevon's paradox and the fungibility of natural gas as compared to that of oil.

## II. DIESEL FUEL VERSUS NATURAL GAS ENERGY ANALYSIS

In order to determine if natural gas can replace diesel for the country's heavy duty truck transportation needs, the energy equivalence and efficiencies of natural gas alternatives should be compared to diesel. There are two alternatives for using natural gas as a replacement for diesel, compressed natural gas (CNG) and liquefied natural gas (LNG). Compressed natural gas is stored at 200 atmospheres of pressure. Even at this pressure CNG has approximately 25% of the energy density of diesel; however, given that it is more knock resistant than diesel, it has a higher efficiency since it can be operated at higher compression ratios. However, the volume required for CNG makes it unsuitable for long range transportation. Compared to LNG it is cheaper but requires more infrastructure. It may be a useful alternative for other heavy duty truck applications that do not require large fuel tanks.

Liquefied natural gas (LNG) has a much greater energy density than CNG. It takes one-third of the volume of CNG. This makes LNG the natural gas alternative of choice for long range transportation, such as coach buses and semi-trucks. Another factor to consider is fuel weight. Although the much lower weight per volume of LNG (3.5lb/gal), compared to that of diesel (7.6lb/gal), more than makes up for the lower energy density of LNG, the complex fuel tank used for LNG storage is much heavier than a fuel tank used for diesel.

In a study sponsored by the U.S. Department of Energy (DOE) and conducted by the National Renewable Energy Laboratory (NREL), a Cummins NTC 350 (350 hp) diesel engine and a Prototype DDC S60G (370 hp) LNG engine were compared to determine how an LNG powered engine performs. Table 1 shows the resulting diesel and LNG transportation performance. These results can be used to determine how much LNG would be needed to supplant diesel in agricultural applications.

TABLE 1
FUEL ECONOMY FOR DIFFERENT VEHICLES

| Vehicle Number | Energy Content(Btu/gal) | Fuel Economy(mi./gal) | Diesel Equivalent Fuel Economy(mi./DEG) |
|---|---|---|---|
| LNG1 | 76,000 | 2.6 | 4.3 |
| LNG2 | 76,000 | 3.0 | 5.1 |
| LNG3 | 76,000 | 2.9 | 4.9 |
| DSL4 | 128,500 | 4.5 | 4.5 |

The resulting average natural gas fuel economy was 2.8 miles per gallon of LNG, which corresponds to 4.7 miles per diesel equivalent (DEG). The ratio 4.7/2.8= 1.67 can be used to determine the total amount of LNG needed, and from that the total cubic feet of natural gas production required to replace diesel, as will be shown below.

### A. Natural Gas Quantities for Fleet Conversion

Under the US department of Transportation is an organization called the Bureau of Transportation Statistics. They keep track of a large repository called the National Transportation Statistics (NTS), which has information pertaining to the transportation sector in the United States. This repository is publicly available online and is updated quarterly. NTS Table 4-5 displays Fuel Consumption by mode of transportation. This table has 7 main categories, Air, Highway, Transit, Rail Freight, Amtrak, Water and pipeline. Our focus is on the Highway category. In this category we will not focus on light duty short wheel base or light duty long wheel base vehicles. The focus will be on Single unit 6 tire, combination trucks, and buses. This second set is most likely to be diesel users. They are also more likely to have fixed or semi fixed routes which would make the transition to natural gas which has fewer fueling stations an easier one. The total number of gallons of fuel used by this group is 46,341 million gallons per year, which account for 27.56% of the total gallons of fuel used in the highway category.

For our calculation we assume all 46,341 million gallons are diesel. This is a fair assumption because the three subcategories are heavy movers which are almost entirely dominated by diesel engines. According to the US Department of Energy's Annual Energy Review 2010 the fuel economy of heavy duty trucks has stayed relatively constant over the last 40 years at a value of around 6 miles per gallon. The fuel economy for heavy duty trucks in 2008 was 6.2 miles per gallon. Using this fuel economy number and the number obtained from the Bureau of Transportation Statistics for gallons of fuel, we can back calculate the number of miles in this category to be 287,314 million miles per year.

LNG has a diesel gallon equivalent (DGE) of 1.67 LNG gallons per every 1 gallon of diesel. This is an energy based derivation. The Office of Energy Efficiency & Renewable Energy commissioned an Advanced Technology Vehicle Evaluation study called the "Norcal Prototype LNG Truck Fleet". The study concluded that the heavy duty trucks using LNG had an average fuel economy of 4.3 miles per DGE. They also found that in the same trucks diesel engines got an average of 4.89 miles per DGE. This means that the LNG trucks have a relative mileage of 88% when compared to their diesel engine counterparts. Since 1 DGE is 1.67 gallons of LNG that translates to about 3.27 miles per gal of LNG. This means that if 287,314 million miles were driven by our target conversion group of vehicles that would have needed 87,918 million gallons of LNG for one year.

A gallon is equivalent to 3.7854 liters. LNG has a density of about .5 kg per liter. This means in one year 166,402 million kilograms of LNG would be needed. LNG is not the normal mechanism by which natural gas is used or transported. Natural gas is usually handled in its gaseous form and measured in cubic feet. For our calculations we will consider natural gas to be 100% methane ($CH_4$). This is not exactly true but will suffice for our calculations. Methane has gaseous density of .717 kg per cubic meter. Using this number we find that 166,402 million kilograms of LNG is about 232,082 million cubic meters of gaseous natural gas. Since there are 35.31 cubic feet per cubic meter this equates to 8.2 trillion cubic feet of natural gas needed for this sector of transportation.

According to the 2010 Annual Energy Review the United States consumed 24.1 trillion cubic feet of natural gas. Of this amount, the United States produced 21.6 trillion cubic feet. If heavy duty trucks were to be converted to run on natural gas in the form of LNG the United States would have consumed about 32.5 trillion cubic feet of natural gas per year. That is 134% of its current consumption. If the United States continued to produce natural gas at current levels, it would need to import 33% of the natural gas it consumes. This is unlikely because the price of natural gas is very low compared to historic prices. This huge increase in consumption would lead to an increase in price and an increase in domestic natural gas production. Further, the import capabilities are not sufficient to sole source the increase from imports. The topic will be discussed more in a later section.

Table 1
On the other hand, the conversion of the heavy duty truck fleet would displace 46,341 million gallons of diesel. A barrel of oil is volumetrically equivalent to 42 gallons. So volumetrically the conversion would displace 1.1 billion barrels of oil. According to the 2010 Annual Energy Review the United States consumed 19.2 million barrels per day or 6.835 billion barrels per year. The displacement of 1.1 billion barrels for that year would have decreased the total oil consumption of the United States to 5.731 billion barrels per year. In 2010, the United States produced 9.4 million barrels per day or 3.346 billion barrels for the year. The United States imported 49% of what it consumed. After the conversion, the United States would have imported only 2.385 billion barrels or 42% of what it consumed.

However, a barrel of crude oil does not produce a barrel of diesel but the other products of the crude oil resold for export.

TABLE 2
LNG NEEDED TO SUPPLANT DIESEL

| Quantities | (in millions) | Conversions | |
|---|---|---|---|
| gallons of diesel | 46,341 | diesel mi/gal | 6.2 |
| Miles | 287,314 | LNG mi/gal | 3.267978 |
| gallons of LNG | 87,918 | L/gal | 3.7854 |
| kg of NG | 166,402 | kg/L of LNG | 0.5 |
| m3 of NG | 232,082 | kg/m3 CH4 | 0.717 |
| ft3 of NG | 8,194,799 | m3/ft3 | 35.31 |

*B. Natural Gas reserves versus current production*

In order to replace diesel for heavy duty truck transportation with a natural gas alternative, the current natural gas production and consumption need to be evaluated to ensure that there is sufficient natural gas for the country's other energy needs. Currently, most of the natural gas that is consumed in the U.S. is produced domestically, with a small amount of net natural gas being imported from Canada. The following table compares the natural gas reserves and resources to the current production. It also shows the production rate or importation increase necessary to support complete replacement of diesel with natural gas for heavy duty truck transportation.

TABLE 3
UNITED STATES NATURAL GAS SUPPLY VS CONSUMPTION

| | Proven Reserves | Recoverable Resources | Total Resources |
|---|---|---|---|
| **Natural Gas (Tcf)** | 272 | 2,744 | 14,000 |
| **Years at current consumption** | 11.29 | 113.86 | 580.91 |
| **Years after fleet conversion** | 8.37 | 84.43 | 430.77 |

Table 3 shows that in the United States there is a very large amount of total natural gas resources amounting to a total of 14,000 trillion cubic feet. Unfortunately only 2,744 trillion cubic feet of this is technically recoverable. Of that technically recoverable portion only 272 trillion cubic feet is proven reserves. As table 3 shows the proven reserves are sufficient for a little over a decade. The recoverable resources on the other hand are sufficient for over 113 years at current consumption rates. The total resources can supply almost 600 years of current consumption. This number is interesting to note but not yet practical to consider because it is not recoverable with current technology.

With a fleet conversion the years of supply change slightly. The US total resources of hydrocarbon resources supply around 430 years which is still fairly large but again not yet practical. The more practical numbers to be considered are the years of available technically recoverable resources and proven reserves. After the fleet conversion the supply of technically recoverable resources lasts for 84 years while the proven reserves only last for about 8 years.

The issue can be addressed via an increase of natural gas production which can transfer years from the resources categories to the proven reserves category. Between the 50's and the mid-80's, natural gas production was at pace with consumption. Afterwards consumption began to outpace production, increasing the need for imports as can be seen in the figure below.

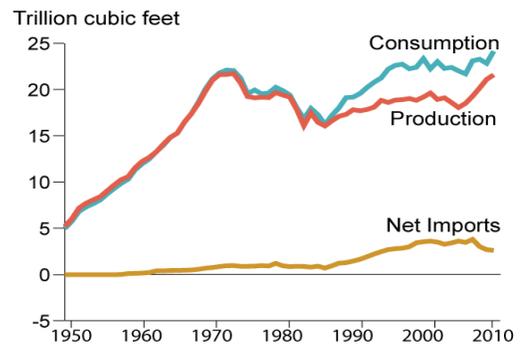

Fig. 1. U.S. Natural gas consumption, production, and net imports, 1949-2010

As can be seen in Fig. 1, in the last few years, production has been increasing at a much faster pace than the increase in consumption, and the net difference has become smaller. This has been made possible by the advent of more efficient drilling techniques and the increased production of natural gas from shale formations. Fig. 2 gives a better interpretation of the role of production of natural gas from shale, and its importance in the increased production capacity needed to meet the requirements to replace diesel for heavy duty truck transportation.

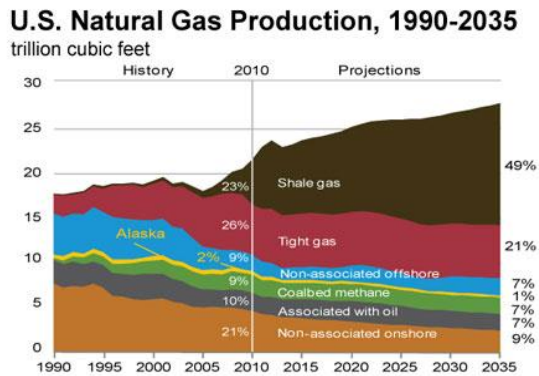

Fig. 2. U.S. Natural gas production, 1999-2035

### III. DOMESTIC TRANSPORTATION CAPACITY EFFECT

The main transportation method for natural gas is via pipeline. LNG distribution is via trucking. According to information from LNG Global Limited, as of April 2012 there is only one LNG liquefaction facility in the US, and it is located in Alaska. It has been supplying LNG to Japan for the last 42 years. The LNG supply in the US comes mainly via imports. A new liquefaction plant at Sabine Pass in Texas is planned but its intended use if is for exportation. This type of domestic supply will not be sufficient for such a large demand for LNG if the heavy duty fleet were to be converted. The first step would be to build a large number of regional liquefaction facilities and distribute LNG regionally by truck. To supply regional liquefaction facilities with natural gas the existing pipeline network would be used.

An added natural gas consumption of 8.2 trillion cubic feet would necessitate a large increase in transport via pipeline. Luckily the transport capacity within the US does not seem to be an issue on the whole. The transport capacity of natural gas in the US is around 67 trillion cubic feet per year. The problem is that the transport capacity is large in some areas and small in others. Already the price of natural gas varies across the United States due to these bottlenecks. The Natural gas price at Henry hub is around $2.50/mmBTU while in the northeast it is around $4/mmBTU. This difference comes from area congestion. So there is a transportation issue currently but that should be negated easily with an increase in capacity. This makes it a short term issue but not a long term one. Even the increase of natural gas consumption due to a fleet change should prove to be only a minor short term issue.

#### A. Carbon Dioxide Reduction Potential

The conversion of the heavy duty fleet has not only been touted as an issue of energy independence but also an issue of de-carbonization of the transportation fleet which arguably will aid in slowing the human contribution to global warming. As mentioned before, the 2010 heavy duty truck consumption of diesel was 46,341 million gallons for the year. Diesel has a density of .9 kilograms per liter. At this density the mass of diesel consumed was 157,877 million kilograms. According to EngineeringToolBox.com the specific carbon dioxide emission from diesel fuel is 3.2 kg of $CO_2$ per kg of diesel fuel. That means that for the 157,877 kilograms of diesel consumed 505,207 million kg of $CO_2$ was produced.

For the converted fleet, 166,402 million kilograms of LNG would have been used. Again, we assume that the natural gas in LNG is pure methane for our calculation. We also assume that the combustion of $CH_4$ is complete. The mass ratio of carbon to $CH_4$ is 75%. This means that the LNG consumed has 124,802 kg of carbon. The mass ratio of $CO_2$ to carbon is 3.67. This means that the carbon in the LNG will produce 457,607 kilograms of carbon dioxide. These results in a decrease of 47,600 kg of $CO_2$ released when compared to the diesel fuel heavy duty fleet, which is a reduction of about 9.4%.

TABLE 3
FLEET CONVERSION EFFECT ON CARBON DIOXIDE EMISSIONS

| Quantities | (in millions) | Conversions | |
|---|---|---|---|
| gallons of diesel | 46,341 | L/gal | 3.7854 |
| kg of diesel | 157,877 | kg/l Diesel | 0.9 |
| kg of $CO_2$ | 505,207 | $kgCO_2$/kg Diesel | 3.2 |
| gallons of LNG | 87,918 | kg/L LNG | 0.5 |
| kg of NG | 166,402 | | |
| kg of $CO_2$ | 457,607 | | |

#### B. Applicability of Jevon's Paradox

The economic principle behind Jevon's paradox is that efficiency creates a relative decrease in price. This relative decrease in price drives more consumption. This argument is usually used to discount efficiency increases as means to decrease carbon dioxide emissions. This principle does not apply to the carbon dioxide emissions decrease seen above because the decrease is brought upon by a fuel change and not an efficiency gain. Further, because of the relatively high cost of LNG currently seen in the United States, the total expenditures on fuel for this converted sector will actually increase with the LNG model even though natural gas prices are very low and diesel prices are considered high.

#### C. Fungibility in a Global Natural Gas Market

The global oil market is considered fungible. This is because significant transport capacity exists in the global market such that prices do not vary widely. The oil markets remain liquid due to these high transport capabilities. For natural gas to become fungible like oil, it too must have highly liquid markets. A good benchmark of the liquidity of a global market is the import/export capabilities in participating countries. If natural gas could have an export/import capability as a percentage of the market similar to that of oil then it too could be a fungible commodity.

In the last 10 years, oil imports went as high as 60% of total US consumption. Total US consumption of natural gas is 24.1 trillion cubic feet per year. That means in order to have the same export/import capability as a percentage of total consumption, natural gas must also have export/import capabilities near 60% of the market. This equates to export/import capabilities of 14.46 trillion cubic feet per year.

LNG is the means by which natural gas is transported internationally to other markets outside North America. That means there would need to be liquefaction & regasification capacity of 13.7 trillion cubic feet per year available for export/import. The Annual Energy Outlook 2012, published by the US Energy Information Agency [3], reports that by 2016 the United States will have an LNG export capacity of 1.1 billion cubic feet per day, which is 4 trillion cubic feet per year. That means that in order to make natural gas a fungible commodity similar to oil the United States would have to more triple that forecasted number in order to get to the 60% market export/import capabilities seen in the oil markets.

TABLE 4
NATURAL GAS AS A FUNGIBLE COMMODITY

| | | |
|---|---|---|
| US max oil import percentage | 60% | |
| US Yearly Consumption of NG | 24,100,000 | millions ft3 |
| 60% of market | 14,460,000 | millions ft3 |
| US LNG Export Capacity 2016 | 4,015,000 | millions ft3 |

IV. CONCLUSION

Natural gas may prove to be a worthy source for fuel. Its ability to displace diesel as a transportation fuel has been considered on the energy basis and on the quantity basis. It has the possibility of displacing a large portion of transportation fuel, but comes with the necessary infrastructure investment. Although it is shown that the proven reserves will last for a decade or less, the resources prove to be much larger and may sustain current consumption for hundreds of years.

Natural gas is touted as a clean technology that can significantly reduce carbon dioxide emissions. However, it was shown that because of LNG's lower relative effectiveness the expected reduction in CO2 emissions is not very significant. This is even before we consider the energy usage to liquefy the natural gas which may end up showing that the conversion is carbon neutral. Lastly, fungibility in the natural gas market may still be very far off due to its relative non-liquidity and due to limited export/import capabilities, as compared to the fungible oil market. For these reasons it seems that natural gas conversion for transportation is not as attractive a picture as advocates propose.